\documentclass[preprint,showpacs,preprintnumbers,amsmath,amssymb,superscriptaddress,floatfix]{revtex4}
\usepackage{latexsym}
\usepackage{epsfig}
\usepackage{graphicx}
\usepackage{textcomp}

\begin{document}
\title{Correlated photon pairs generated from a warm atomic ensemble}%
\date{\today}%
\author{R. T. Willis}

\affiliation{Joint Quantum Institute, Department of Physics, University of Maryland and National Institute of Standards and Technology, College Park, Maryland 20742, USA}
\author{F. E. Becerra}
\affiliation{Joint Quantum Institute, Department of Physics, University of Maryland and National Institute of Standards and Technology, College Park, Maryland 20742, USA}
\affiliation{Departamento de F\'{\i}sica, CINVESTAV, Apdo. Post. 14-740, 07000 M{\'{e}}xico, Distrito Federal, M{\'e}xico}
\author{L. A. Orozco}
\affiliation{Joint Quantum Institute, Department of Physics, University of Maryland and National Institute of Standards and Technology, College Park, Maryland 20742, USA}
\author{S. L. Rolston}
\affiliation{Joint Quantum Institute, Department of Physics, University of Maryland and National Institute of Standards and Technology, College Park, Maryland 20742, USA}

\begin{abstract}
We present measurements of the cross-correlation function of  photon pairs at 780 nm and 1367 nm, generated in a hot rubidium vapor cell.   The temporal character of the biphoton is determined by the dispersive properties of the medium where the pair generation takes place.  We show that short correlation times occur for optically thick samples, which can be understood in terms of off-resonant pair generation.  By modifying the linear response of the sample, we produce near-resonant photon pairs, which could in principle be used for entanglement distribution.
\end{abstract}

\maketitle

\section{Introduction}
Entangled pairs of photons are currently the leading option for achieving long-distance secure quantum communication but suffer from loss in fiber optic channels.  Duan \textit{et al.} \cite{duan01} suggested a method for overcoming this by inserting atomic ensemble-based quantum repeaters in the communication channel.  In response, significant experimental and theoretical effort has gone into understanding and demonstrating quantum light-matter interfaces \cite{eisaman04,eisaman05,chou05,chaneliere05,sherson06}.  An integral part of many of the schemes based on the ideas of Ref. \cite{duan01} is a source of correlated photon pairs that are resonant with atomic quantum memories.  Spontaneous four-wave mixing (FWM) in atomic ensembles is a well-suited candidate for this application \cite{balic05,polyakov04,kuzmich03}, complimenting the work done in optical fibers \cite{Li05} and in non-linear crystals \cite{neergaard07}.  Chaneliere \textit{et al.} \cite{chaneliere06} proposed a quantum repeater that operates at telecommunication wavelengths and experimentally demonstrated pair production from a laser-cooled, cold atomic ensemble with one photon in the telecommunications band and the other in the near-IR.  Balic \textit{et al.} \cite{balic05} and Du \textit{et al.} \cite{du08} have demonstrated biphoton generation in the near-IR with controllable correlation times.

Here we demonstrate a source of correlated photon pairs, or biphotons, with one photon at 780 nm and the other at 1367 nm.  The photons are created in a laser-pumped warm atomic ensemble (isotopically pure $^{85}$Rb) using a four-wave mixing (FWM) interaction that is resonantly enhanced by the diamond energy level structure \cite{becerra08,willis09}  shown in Fig. \ref{struct}.  The 780-nm wavelength photon is compatible with atomic quantum memories \cite{chaneliere05}.  The 1367-nm photon is near the zero-dispersion telecommunication window in standard optical fibers.
For quantum information science applications, it is important to consider the spectral and temporal properties of the biphotons as well as the quantum mechanical correlation properties of the pairs.

We will present results on the quantum correlation and entanglement of the pairs in a subsequent article and here will focus on the spectral and temporal character of the photon pairs as measured by cross correlations. We find that the gain of the non-linear FWM interaction and the linear absorption/dispersion of the medium determine the biphoton wavepacket. Changing the parameters of the pump lasers can modify both the gain and the absorption/dispersion of the medium. Spectral filtering and additional laser fields modify the biphoton correlation time and frequency.

The work in Refs. \cite{freedman72,aspect81,aspect82} has shown that correlated photon pairs can be generated from ladder-level structure atoms.  In these schemes the photons are emitted uniformly in all directions, leading to a very low rate of coincidence on two detectors, which capture a very small portion of the solid angle.  Biphotons created by non-linear, phase-matched processes such as parametric down conversion in non-linear crystals \cite{kwiat95} or spontaneous FWM are highly correlated in direction.  The enhancement in correlations in the present system system can be understood using a perturbative stepwise description of the generation process.

We describe in the Section II  our experiment and show the results.  In section III we discuss a theoretical description that captures many of the features of the data.  We also discuss how the thermal motion of the atoms contributes to the correlation function.  In a final concluding section IV we review our results.

\section{Experiment}

\begin{figure}
\includegraphics[width=3.1in]{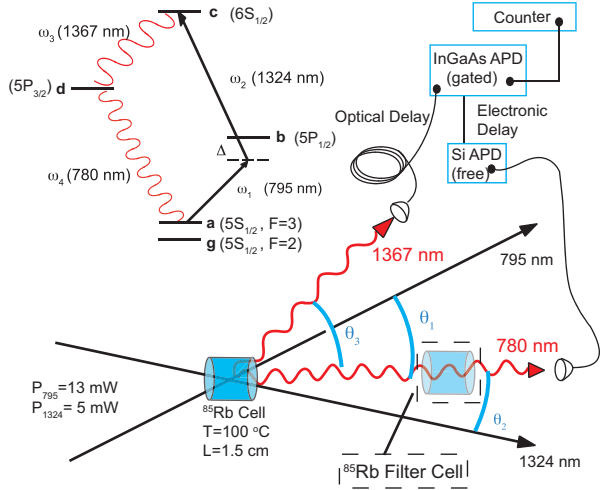}
\caption{\label{struct}(Color Online)  A schematic of the experimental apparatus and a simplified energy level structure. The angles defining the beam directions are $\theta_1$= 2.0$^\circ$, $\theta_2$= 0.7$^\circ$, and $\theta_3$= 2.7$^\circ$. The Rb filter cell can be moved in and out of the beam path of the 780-nm photon.}
\end{figure}

Figure \ref{struct} shows the relevant atomic structure for our experiment as well as the geometry of the applied and generated light fields.  We apply two pump lasers continuously to a 1.5-cm long isotopically pure $^{85}$Rb cell held at 100 $^\circ$C.   The 795-nm ($\omega_1$) and 1324-nm  ($\omega_2$) pumps interact with the atoms to generate pairs of photons at 780 nm ($\omega_3$) and 1367 nm ($\omega_4$).  The correlated photon pairs fulfill phase matching and energy conservation as they come from a FWM process.  Correlated pairs are generated on opposite sides of two concentric cones; we collect pairs that are co-planar with the pump beams.  A short single-mode fiber delivers the  780-nm photon  to a silicon avalanche photodiode (APD).  The 1367-nm photon travels through a 200-m long single mode fiber, which delays it by $\approx$ 1 $\mu$s and then delivers it onto a InGaAs APD which is gated on for 1-ns intervals.   The silicon and InGaAs APDs have detection efficiencies of 40\% and 20\% respectively.  Preceding each single mode fiber is a suitable interference filter and polarization analyzer.  Both pump lasers are horizontally polarized.  We measure cross-correlation functions by first recording a count on the 780-nm APD and then gating on the 1367-nm APD with variable delay.

We align the system by first directing all three pump beams necessary for FWM on the ensemble and
maximize the generated light. We then optimize the coupling of the 1367 nm light into its mode fiber with an efficiency of up to 70\%. We also couple the 780 nm pump beam into its single mode fiber with a maximum efficiency of 80\%. Then, in a  final alignment step, we remove the 780 nm pump and observe and maximize the pair coincidence rate.

We consider two distinct pumping regimes.  In the first case the lower pump laser (795-nm) is detuned from intermediate state resonance ( off-resonant pumping).  In the second case (resonant pumping)  the lower laser is tuned to be on intermediate resonance.  These two pump-laser configurations lead to significant differences in the biphoton properties.

\subsection{Off-resonant pumping}
\begin{figure}
\includegraphics[width=3.1in]{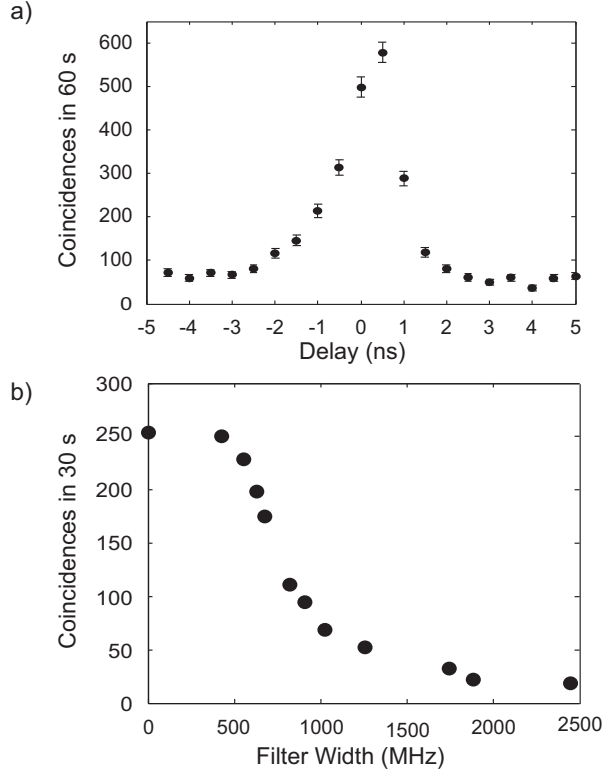}
\caption{\label{offres}Experimental results of: (a) The un-normalized cross-correlation function for a 1.5 GHz red-detuned 795-nm pump and no filter cell.  (b) Coincidence counts at zero delay as a function of the filter cell  spectral width. The zero corresponds to the resonance of the 5S$_{1/2}$ F=3 $\rightarrow$ 5P$_{3/2}$ F=4 transition.}
\end{figure}

Figure \ref{offres}a shows a typical measured cross-correlation function with the 795-nm pump laser detuned $\Delta=1.5$ GHz below the 5S$_{1/2}$ F=3 $\rightarrow$ 5P$_{1/2}$ F=2 transition, well outside the Doppler width.  The 1324-nm pump laser frequency closes the  5S$_{1/2}$ F=3 $\rightarrow$ 6S$_{1/2}$ F=3 two-photon transition in the level scheme of Fig.~\ref{struct}.  The polarization analyzers for the signal beams pass vertically polarized photons.
The width of the correlation function is short ($\approx$ 1 to 2 ns), about the resolution of the detector, rather than of order the lifetime of the intermediate $P_{3/2}$ state as might be expected.
To understand the temporal and spectral character of the correlated pairs, we insert a second hot isotopically pure $^{85}$Rb cell into the path of the 780-nm photons as illustrated in Fig. \ref{struct}  to act as a narrow-band filter. We record coincidence counts at zero delay and vary the temperature of the filter cell, changing the width of the transmission region.  Fig. \ref{offres}b shows the coincidences in 30 s as a function of the filter width  (defined as 50\% transmission).   No change in the coincidence rate occurs until the filter cell absorption is wider than that of the FWM cell, implying that
the correlated 780-nm photon of the pair is far detuned from the intermediate state resonance.

The FWM-cell optical depth for resonant 780-nm photons  can be as high as 20 at our operating temperature; the biphotons generated near resonance (as is expected) are re-scattered  and do not reach the detectors in the phase-matched direction.  The photons that emerge from the cell are detuned from the intermediate resonance (level d in Fig. \ref{struct}) and this gives rise to the short correlation time.  The measurement of the transmission as a function of filter width gives a bandwidth of the 780-nm component of the biphoton of 350 MHz.  Note this is in contrast to Ref. \cite{chaneliere06}, which observed long ($>15$ ns) correlation times from a laser-cooled sample.  In this case the absence of Doppler broadening allows photons much closer to resonance to be transmitted, and thus longer correlation times. We note that this effect provides an alternative explanation to the observed shortening of the correlation times as the optical depth is increased, attributed to superradiance in Ref. \cite{chaneliere06}.

\subsection{Resonant pumping}
\begin{figure}[ht]
\includegraphics[width=3.1in]{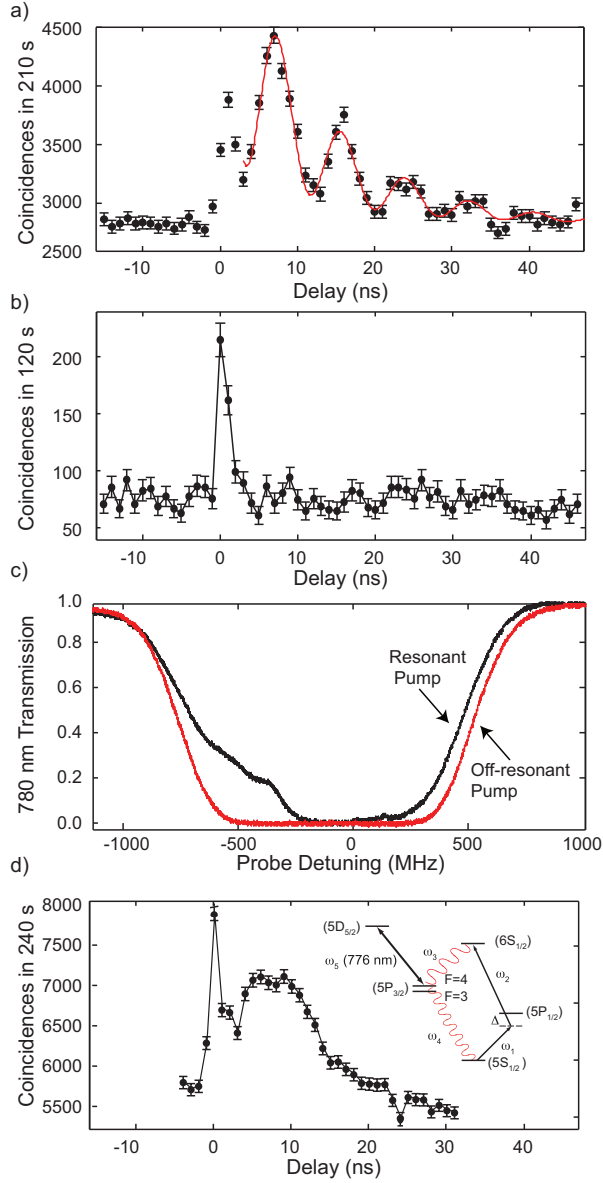}
\caption{\label{onres}(Color Online) Experimental results of: (a) Cross-correlation function with a resonant 795-nm pump beam,  the red curve is a decaying oscillation at the hyperfine interval. (b)  The cross-correlation function with a $^{85}$Rb filter cell (same temperature as FWM cell)  in the 780-nm beam path.  (c)  The transmission of a weak 780-nm probe beam in the FWM cell and the FWM cell with resonant 795-nm pump laser. (d) Cross-correlation as in (a) except that  a laser at 776 nm couples one of the hyperfine levels of the $5P_{3/2}$ state out of the diamond as shown in the inset.  Note the suppressed zeroes in the coincidence counts in (a), (b), and (d). The solid line in (b) and (d) is a guide for the eye.}
\end{figure}

If we vary the frequency of the pump lasers, the cross-correlation function changes. With the 795-nm pump laser set close to the 5S$_{1/2}$ F=3 $\rightarrow$ 5P$_{1/2}$ F=2 transition and the 1324-nm pump laser 100 MHz above of the two-photon, zero-velocity class resonance (see Fig. \ref{onres}a), the correlation function shows two distinct features. The narrow peak at zero delay is similar to the one seen in the off-resonant pump case (Fig. \ref{offres}), while the second feature shows a longer decay with oscillations, attributable to quantum beats.

The Fourier transform of the correlation shows a peak at approximately 120 MHz, consistent with the hyperfine splitting between the $F=3$ and $F=4$ of the $5P_{3/2}$ $^{85}$Rb excited state.  Multiple decay paths exist due to the hyperfine splitting of the intermediate state \cite{aspect84,chaneliere06}. The choice of the $F$ sublevel in the initial state allows decay through a different intermediate path that shows different interval frequencies. Additional measurements with a $^{87}$Rb atomic sample also show quantum beats at the appropriate hyperfine intervals of that isotope. The continuous (red) line in Fig.~\ref{onres}a) is a fit to the data from 3-45 ns to an exponentially decaying sinusoid with a frequency equal to the relevant hyperfine splitting of $f=120.6$ MHz.  We find a decay constant of $\tau=11.6 (2)$ ns.

Strong optical pumping due to the on-resonant 795-nm pump beam can change the population between the two hyperfine ground state levels, including modifying particular velocity classes of atoms. Fig. \ref{onres}(c) shows the transmission of a weak 780-nm probe beam through the FWM cell with and without the resonant 795 nm pump beam. There is a clear modification of the FWM cell transmission of 780-nm light due to optical pumping. The specific shape of the  curve  can vary with small changes in the detunings and polarization.  Using the filter cell (in dashed lines on Fig. \ref{struct}) we find that the narrow structure observed in Fig. \ref{onres}(a) arises from far off-resonant 780-nm photons, while the longer time structure occurs due to near-resonant photons transmitted due to a modified absorption profile (Fig. \ref{onres}(b)).

Figure \ref{onres}(d) shows the cross correlation in the presence of an extra laser field resonant with the  $F=4$ $5P_{3/2}$ hyperfine level to the $5D_{5/2}$ manifold \cite{becerra09} in counter-propagating configuration with the FWM beams ( it makes an angle of less than one degree with respect to the 780 nm path).  The additional laser produces a visible electromagnetic induced transparency (EIT) window on the 780 nm transmission when the atomic absorption is large \cite{gea95}, but the effect is no longer visible when we operate the cell at the FWM temperature with the strong 795 nm laser also present. However; under the operational experimental conditions, we do not observe any
modification of the cross-correlation function in the presence of the coupling laser until
its frequency matches the transition frequency of $5P_{3/2}, F = 4$ to $5D_{5/2}, F = 5$, when it suppresses the hyperfine quantum beat oscillations.  The 776 nm
laser on resonance couples to the hyperfine state $F = 4$ and decouples it form the
FWM process. The FWM process in the presence of the 776 nm laser has a single
hyperfine level available in the $5P_{3/2}$ erasing the quantum interference.
The laser provides some degree of optical control of the biphoton wavefunction through the removal of hyperfine states available for decay.

\section{Theory and Analysis}
Consider a gas of atoms with the level structure indicated in Fig. \ref{struct}, all initially in ground state $a$.  Applying the two pump beams with frequencies $\omega_1$ and $\omega_2$ and k-vectors $\vec{k}_1$ and $\vec{k}_2$ to the atoms can lead to an excitation of a single atom to upper excited state $c$.  A spontaneously emitted 1367-nm photon, with frequency $\omega_3$ and k-vector $\vec{k}_3$, and can then be emitted into all directions with equal probability leaving one atom in the ensemble in the excited state $d$.  It is impossible to know which atom did the scattering and so the system is left in a state of one delocalized atomic excitation shared by all the atoms.  Once the 1367-nm photon is emitted, a spatially varying phase is imprinted on the collective atomic state.  Ideally the state can be written as
\begin{equation}
\label{spinwave}
\left|\psi\right>=\frac{1}{\sqrt{N}}\sum_{j=1}^{N}e^{i\vec{k}_4\cdot\vec{r}_j}\left|d_j\;a_{\mathrm{else}}\right>\;.
\end{equation}
Here $N$ is the number of atoms, $\vec{r}_j$ is the position of the $j$th atom, and the ket is short hand for the state in which the $j$th atom is excited with all other atoms in the ground state.  The wave vector $\vec{k}_4$ is given by $\vec{k}_4=\vec{k}_1+\vec{k}_2-\vec{k}_3$.  In the case that the 1367-nm photon was emitted into the phase-matched direction, upon the decay of the state in Eq. \ref{spinwave}, the 780-nm photon is emitted preferentially in the $\vec{k}_4$ direction.   The interference gives rise to an enhancement in the correlations of the two photons.

We present here a qualitative model to describe the cross-correlation function.  A full theoretical treatment of pair production from an atomic ensemble is quite involved.  The generated fields must be treated quantum mechanically and both the linear absorption and non-linear gain of the fields must be included.  The beam geometry and the thermal motion of the atoms also contribute to the generated biphoton properties.  The approach we take here is to build the simplest model that gives qualitative agreement with the data.  We show that the temporal profile of the emitted pairs is well described by taking into account the resonance structure of the emitting transitions and the frequency dependent attenuation of the photons in the generation medium.  More complicated treatments are possible; References \cite{lukin00,kolchin07} present models with propagation which incorporate the parametric gain and the linear response of the medium in similar configurations.

\begin{figure}[ht]
\includegraphics[width=3.1in]{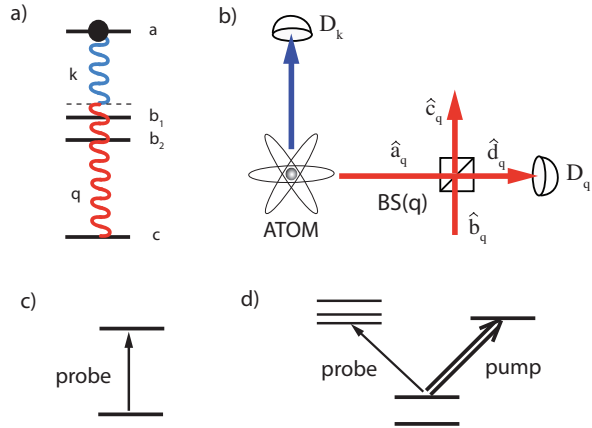}
\caption{\label{theorydiagram}(Color Online) a) Level structure for the atom which creates the light field used in the model.  b) Schematic of the model (see text).  c) Depiction of a the two-level atom used to create the filter function in the off-resonant pumping case.  d)  Level structure used to create the filter function in the on resonant case.}
\end{figure}

Consider a single four-level atom with the level structure shown in Fig. \ref{theorydiagram}a, initially in the excited state.  The atom decays and emits two photons in the process.  Photon $k$ ($q$) is emitted on the upper (lower) transition with frequency $\nu_k$ ($\nu_q$).  Two non-degenerate intermediate states provide two possible decay paths leading to quantum beats.  The two photons are incident on two separate single-mode detectors $D_k$ and $D_q$.  Before arriving on the detector, photon $q$ is incident on a beamsplitter (BS) with a frequency-dependent transmission coefficient $t(\nu_q)$.  The un-normalized cross-correlation (CCF) function can then be
written as
\begin{equation}
\label{coinrate}
\left|\Psi\right|^2=\left|\left<0|\hat{E}^+_{D_k}(t_1)\hat{E}^+_{D_q}(t_2)|\psi\right>\right|^2\;.
\end{equation}
Here, $\hat{E}^+_{D_k}(t)$ and $\hat{E}^+_{D_q}(t)$ are the positive-frequency field operators at the two detectors \cite{mandel_wolf}.  The ket $\left|\psi\right>$ is the two-photon quantum state of the field from the decaying atom which can be calculated using the Wigner-Weiskoff appoximation \cite{scullybook} and can be written as
\begin{equation}
\label{fieldstate}
\left|\psi\right>=\sum_{k,q,j}-g^j_{a,k}g^j_{b,q}L_{kq}L^j_q\left|1_k,1_q\right>\;,
\end{equation}
where $L_{k,q}=\left[i(\nu_k+\nu_q-\omega_{ac})-\Gamma_a\right]^{-1}$ and $L^j_q=\left[i(\nu_q-\omega^j_{bc})-\Gamma_b\right]^{-1}$ and we have assumed the atom is at r=0.  The sum over $j$ in Eq. \ref{fieldstate} is over the two possible decay paths.  The coupling constants are given by $g^j_k=-\vec{\mu}\cdot\hat{\epsilon}_k\mathcal{E}_k/\hbar$ where $\mathcal{E}_k=\sqrt{{\hbar\nu_k}/{2\epsilon_oV}}$ with $V$ the appropriate mode volume.

The fields before and after the frequency-dependent BS are related by the quantum mechanical BS relations.  Following the convention in Fig. \ref{theorydiagram}b $\hat{d}_q=t(q)\hat{a}_q+r(q)\hat{b}_q$, where $r(q)$ is the field reflection coefficient.  In our model mode $\hat{b}_q$ is vacuum and does not contribute to the calculation of cross-correlation function.  The result is that the field operator for detector $D_q$ is modified by $\hat{E}^+_{D_q}(t)=\int \mathcal{T}(t-\tau)\hat{E}^+_{D_q^{\mathrm{free}}}(\tau)d\tau$ where $\hat{E}^+_{D_q^{\mathrm{free}}}(t)$ is the free-space field operator and $\mathcal{T}(t)$ is the time-domain representation of the transmission function of the BS.  This treatment is the same as given by Ref. \cite{eberly77}.

The interpretation of the model, as it relates to our experimental system, is as follows:  The photon pair emitted by the single excited atom is analogous to a photon pair created via spontaneous FWM somewhere in the atomic ensemble.  The lower photon, at 780 nm in the experiment and labelled $q$ in the model, must propagate out of the medium, which has a transmission coefficient $t(q)$.  We consider two different transmission functions: one related to the off-resonance pumping case and the other related to the on-resonance situation.

\subsection{Off-Resonance Case:  Two level atom filter}
\begin{figure}[ht]
\includegraphics[width=3.1in]{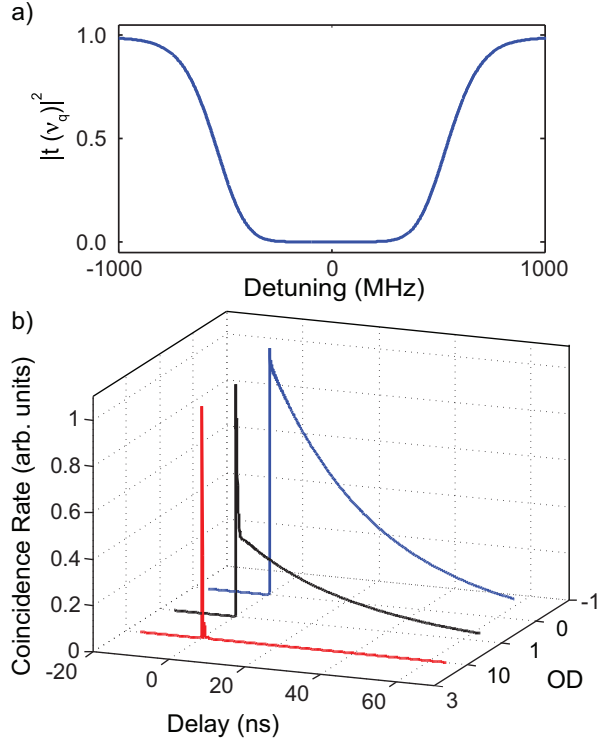}
\caption{\label{offrestheory}Calculation of: a) The transmission using the model of the frequency dependent beamsplitter for an optical depth of 10 as a function of detuning. b) Theoretical prediction for coincidence rates for the off-resonantly driven case at three different optical depths (the axis is logarithmic).  }
\end{figure}

Figure \ref{onres}(c) shows the transmission of a weak 780 nm probe beam for off-resonance pumping (red line).  This transmission is well approximated by that of a collection of Doppler broadened two level atoms.  Specifically we use $t(\nu_q)=e^{i\alpha\bar{\chi}(\nu_q)}$ where $\alpha$ is a real constant involving the density and length of the sample and $\bar{\chi}(\nu_q)$ is the Doppler-broadened linear-response function for the filter atoms.  Before Doppler averaging, the response function for a single velocity-class ($v=0$) of atoms is $\chi(\nu_q)=\left[i(\nu_q-\omega_f)-\Gamma_f\right]^{-1}$.  Fig. \ref{offrestheory}(a) shows the modulus squared of the transmission function for maximum optical depth of 10; it shows similar behavior to the experimental case of Fig. \ref{onres}(c).  Fig. \ref{offrestheory}(b) shows the expected correlation function for three different OD values.  The temporal extent of the correlation function becomes narrower at high OD as the spectral content of the 780-nm photon that contribute to measured coincidences is pushed further from resonance.  At very low OD we see the full natural lifetime decay constant associated with the intermediate level of the source atoms.  At ODs of more than 10 the CCF is very narrow with an extent of less than 1 ns.  In the experiment the observed CCF is $\approx$ 1 ns wide, which is about the absolute resolution of our detection system.  Decreasing the OD should lead to a longer decay time, however at low enough optical depths to detect a significant change in the temporal profile, the coincidence rate is too low to acquire adequate data.

\subsection{On-Resonance Case:  Strongly Driven Vee}
The absorption of the 780-nm photons is strongly modified by the presence of a strong, resonant 795-nm pump beam as is evident in the data in Fig. \ref{onres}c.  To model this type of transmission we turn to a more complicated atomic level structure.  We calculate the response function for a weak probe propagating in an ensemble of atoms with the level structure shown in Fig. \ref{theorydiagram}~d.  This structure is very similar to that of real Rubidium with two stable ground state manifolds as well as multiple levels separated by the appropriate hyperfine splittings in the excited state.  The transmission coefficient is calculated by solving for the steady state response of the system exactly, when only the strong pump is present, then solving pertubatively to lowest order for the probe field.

\begin{figure}[ht]
\includegraphics[width=3.1in]{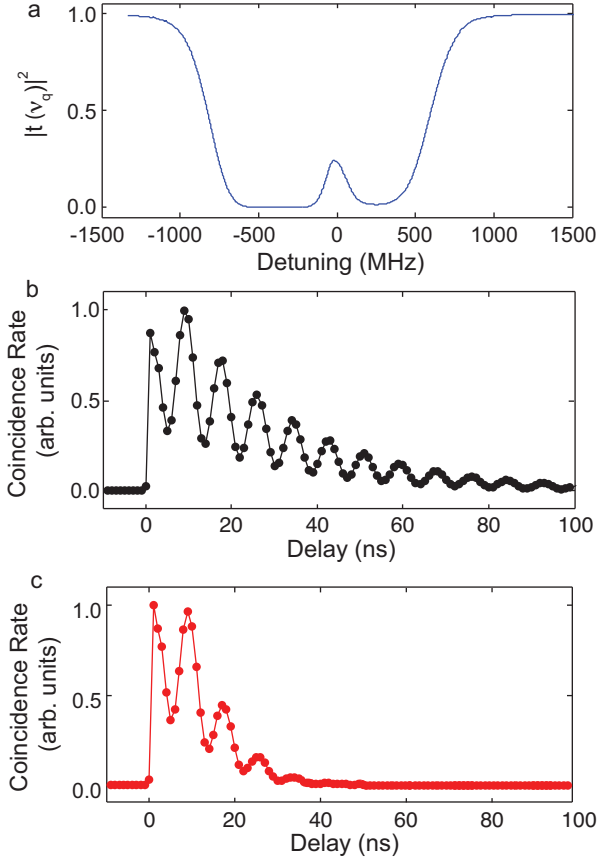}
\caption{\label{onrestheory}Calculation of: a) The transmission function of a collection of atoms with the level structure shown in Fig. \ref{theorydiagram} d.  b)  The cross-correlation function for a strongly resonant pump laser.  c) The cross correlation when the thermal motion of the atoms is taken into account.}
\end{figure}

The transmission can be written as $t(\nu_q)=e^{i\alpha\bar{\chi}(\nu_q)}$ but with a much more complicated expression for $\bar\chi(\nu_q)$.  We do not write the function explicitly but show the transmission with a peak OD of 10 in Fig. \ref{onrestheory}a.  A wide transparency window exists in the response.  Figure \ref{onrestheory}b shows the cross correlation function in this case.  We have binned the results of the calculation to take into account the 1 ns resolution of the detection system.  A peak exists at zero delay as does an exponentially decaying oscillatory piece.  The oscillation frequency matches the splitting between levels $b_1$ and $b_2$ which we have chosen to be the hyperfine splitting of the $F=3$ to $F=4$ levels in the $P_{3/2}$ manifold of $^{85}$Rb.  The decay of the correlation function corresponds to the natural line width of the intermediate state, which here is set to be 6 MHz.  Comparing Fig. \ref{onrestheory}b to the data in Fig. \ref{onres}a it is clear that the decay time of the correlation function is described by the model qualitatively.

\subsection{The Decay of the Correlations}
We now turn to the decay time of the cross-correlation function shown in Fig. \ref{onres}(a) and (d).  Fitting a decaying oscillatory trial function of the form $y_o+\left(A_1+A_2\sin^2(\pi f t+\phi)\right)e^{-t/\tau}$ to the data in Fig. \ref{onres}a gives a decay constant of $\tau\approx12$ ns, less than half that of the intermediate state life-time of 26 ns.  Our theoretical model predicts natural lifetime decays.  In Ref. \cite {chaneliere06} the authors observe a cross correlation that shortens with increasing optical depth and attribute this behavior to super-radiance.  In a double-$\Lambda$ configuration \cite{du08} the cross correlation can be as long one $\mu$s due to slow light effects, which can be tuned by increasing the OD of the sample. This configuration is significantly less affected by Doppler effects.  We measure the correlation function at different atomic densities spanning two orders of magnitude, and find no change in the decay constant of the correlation function, in stark contrast to cold atom experiments.

Our recent work on the time response of FWM to step excitation shows decoherence and decay times faster than the atomic lifetimes involved \cite{becerra10}. The decay time does not depend on the temperature of the cell between 340 K and 380 K, a change of one order of magnitude in the atomic density. The physical quantity of relevance is the induced atomic polarization of the sample, which is affected by Doppler broadening. A density matrix semiclassical formulation that includes Doppler effects as well as absorption and dispersion explains  the results, and they are in agreement with the qualitative model what we present here.

An explanation of the short unvarying decay time is the thermal motion of the atoms.  The two pump lasers excite only those atoms that are moving with  velocities such that the Doppler shift is on the order of a natural linewidth.  For our case  the spread of velocities should be   $\Delta v \approx \Gamma_{2\gamma}/k_{2\gamma}\approx5\;\mathrm{MHz}\times 500\;\mathrm{nm}=$ 2.5 m/s.  Here, $\Gamma_{2\gamma}$ and $k_{2\gamma}$ are the the two-photon linewidth and wavevector, respectively.  According to the description of the generation process, enhanced correlations are due to a spatial phase grating written on the collective excited state (see Eq. \ref{spinwave}).  If the atoms are moving with velocities of a few m/s the grating begins to disappear rapidly once the atoms move a fraction of an optical wavelength.

Consider $N$ spatially phased classical dipole oscillators, which are perfectly phased at time t=0 to interfere constructively in the forward direction.  If those oscillators are then allowed to move ballistically with a Maxwell-Boltzman distribution with thermal velocity $v_t$, the intensity of the forward scattering has the form $I(t)\propto N+(N^2-N)\exp\left[-\tfrac{1}{2}(kv_tt)^2\right]$.  This form of exponential decay also applies to the forward emission of the 780-nm photon from the state in Eqs. \ref{spinwave} and we expect a suppression of the correlation function with the functional form $\exp\left[-\tfrac{1}{2}(kv_tt)^2\right]$.  We fit the data with the assumption that the correlation decays with a natural-linewidth decay time multiplied by an additional motional suppression factor.  We find that the fit is as good in the reduced $\chi^2$ sense as the fit without motional suppression.  The thermal velocity from the fit for the data in Fig. \ref{onres}c is $v_t=6.6$ m/s.  Figure \ref{onrestheory}(c) shows a calculated correlation function when we include the motional suppression.  With this modification, the model is in good qualitative agreement with the data.

\section{Concluding Remarks}
We note that in \cite{du08} and \cite{du08opt} the authors observe a similar narrow feature to what we observe in Fig. \ref{onres}(a), which they attribute  phenomena of the Sommerfeld-Brillouin precursor \cite{oughstun94} \cite{jeong06} at the single-photon level. We have experimentally shown that the broad-band narrow feature arises from photons detuned from the intermediate state.  This provides a simple and direct explanation for the existence a sharp peak in the biphoton wavepacket.

We have measured the structure of the time correlations on the spontaneously generated phase-matched photons from a hot atomic ensemble. We explored different regimes of detuning and showed that the temporal profile is determined in  large part to the dispersive and absorptive nature of the generation medium. We can control the spectral profile of the biphoton cross-correlation with a narrow band optical filter. We can also modify the temporal response by coupling a control laser that addresses a specific hyperfine transition. The biphoton production rate may be increased by orders of magnitude if the absorption of the 780-nm photons could be suppressed.   One possibility would be to use a $\Lambda$ EIT control laser to open a transparency window, but the effect of the EIT laser field on the FWM process will need to be investigated.

The authors would like to thank the group of Alan Migdall at NIST for their loan of the InGaAs APD.  This work was supported by the NSF, DURIP, and CONACYT.

\end{document}